\documentclass[aps,prx,10pt,twocolumn,superscriptaddress,longbibliography,nobibnotes,floatfix]{revtex4-2}

\usepackage{graphicx}
\usepackage{dcolumn}
\usepackage{bm}
\usepackage{lineno}
\usepackage{amsmath}
\usepackage{placeins}
\usepackage{xcolor}
\usepackage{multirow}
\usepackage{makecell}
\usepackage{amssymb}
\usepackage{booktabs, multirow, siunitx}
\usepackage[normalem]{ulem}
\usepackage[colorlinks=true, allcolors=blue]{hyperref}

\usepackage[shortlabels]{enumitem}
\newcommand{\ECNU}{Quantum Institute for Light and Atoms, School of Physics, East China Normal University, Shanghai 200062, China}
\newcommand{\CHONG}{Chongqing Institute of East China Normal University, Chongqing 401120, China}
\newcommand{\SBH}{Shanghai Branch, Hefei National Laboratory, Shanghai 201315, China}
\newcommand{\SHAN}{Collaborative Innovation Center of Extreme Optics, Shanxi University, Taiyuan 030006, China}

\begin{document}
\title{Intelligent Optimal Control of Rydberg Gates with Incremental-Update Deep Reinforcement Learning}

\author{Yue Cai}
\affiliation{\ECNU}

\author{Hanlin Zhang}
\affiliation{\ECNU}

\author{Keye Zhang}
\email{kyzhang@phy.ecnu.edu.cn}
\affiliation{\ECNU}
\affiliation{\SBH}
\affiliation{\SHAN}

\author{Jing Qian}
\email{jqian1982@gmail.com}
\affiliation{\ECNU}
\affiliation{\SHAN}
\affiliation{\CHONG}

\begin{abstract}

Deep reinforcement learning (DRL), acting as a novel and powerful paradigm for quantum optimal control, offers transformative opportunities for advancing neutral-atom quantum computing. In this work, we theoretically demonstrate a DRL-based framework for realizing Rydberg controlled-NOT gates that achieve both high speed and high fidelity through the synchronous modulation of multiple pulse parameters without any prior heuristic ansatz. By introducing an incremental-update learning policy, our framework effectively regularizes the exploration of the control landscape, ensuring the generation of smooth, experimentally feasible pulse profiles while significantly reducing computational overhead compared to conventional schemes. Crucially, the framework autonomously discovers an early-cutoff policy by optimally reconciling operation speed with high-precision coherent control. Our optimized protocol achieves a peak average fidelity of 0.9991, significantly outperforming conventional methods and surpassing the critical fault-tolerant threshold. This work establishes a generalizable, AI-driven pathway for designing high-performance quantum gates and provides a robust paradigm for autonomous control field optimization across diverse qubit platforms.
\end{abstract}

\maketitle

\section{Introduction}

Quantum optimal control (QOC) has emerged as a mainstream approach for quantum gate design in the area of neutral-atom quantum computing \cite{QOC_1,neutral_atom_QC_5,neutral_atom_QC_4,QOC_3,experimental_parameters,neutral_atom_QC_3,neutral_atom_QC_1,QOC_4}. By precisely engineering the control pulses, this technique enables the realization of high-performance quantum gates with improved robustness \cite{QOC_robust_1,QOC_robust_2,QOC_robust_3,QOC_robust_4} or minimal operation time \cite{QOC_time_optimal_2,QOC_time_optimal_1,QOC_time_optimal_3,QOC_fast}. Within the framework of QOC, the pulse-design problem can be formulated as an optimization task, in which an appropriate control policy is imperative by minimizing the cost function aligned with specific optimization objectives. 
Conventional QOC methods such as GRAPE which is widely adopted in pulse optimization \cite{GRAPE,GRAPE_2,GRAPE_3,GRAPE_4}, 
depend on an accurate system modeling and reliable gradient evaluation, making them challenging to apply for high-dimensional, complex quantum systems \cite{GRAPE_defect,GRAPE_defect_2,GRAPE_defect_3}. An alternative way using heuristic search such as the genetic algorithm, demands a prior guess for the pulse shape and thus reduces the effective control freedom, often yielding locally optimal solutions \cite{evolution_algorithm_1,evolution_algorithm_2,evolution_algorithm_4,evolution_algorithm_5,evolution_algorithm_6}. These bottlenecks in both efficiency and effectiveness lead to the urgent need for developing novel QOC methods, particularly for the design of high-fidelity quantum gates in platforms such as Rydberg atom arrays \cite{rydberg_atom_array_1,rydberg_atom_array_2,rydberg_atom_array_3}, where complex many-body interactions and intricate atomic energy-level structures demand sophisticated control strategies.

Deep reinforcement learning (DRL), which integrates artificial neural networks with reinforcement learning, provides a powerful and intelligent framework for addressing challenges in QOC tasks \cite{DRL_1,DRL_2,DRL_3,DRL_4,DRL_6,DRL_2021PRA}.
In this paradigm, an agent interacts with a predefined environment and iteratively refines its behavior through trial and error, where correct actions receive rewards and incorrect actions incur penalties. Through this process, the agent can gradually learn an adaptive control strategy that maximizes the long-term cumulative reward. Operating without access to the underlying dynamical model, its gradients, or an initial guess, all of which are essential to traditional QOC schemes, DRL is intrinsically free from systematic modeling errors \cite{DRL_model_free_1}. This model-free nature makes it a compelling alternative to traditional simulation-based approaches and particularly suitable for a range of quantum control tasks that can be framed as sequential decision-making problems, including quantum circuit design \cite{DRL_circuit_design}, complex target-state preparation \cite{DRL_state_prepaprtion_2,DRL_state_prepaprtion_1}, and suppression of decoherence \cite{DRL_suppress_decoherence}.

In this work, we propose a DRL-based pulse-design framework to address the fidelity-speed trade-off in Rydberg quantum gates, with a specific focus on the challenging native controlled-NOT (CNOT) operations \cite{EIT_CNOT_1,CNOT_2,CNOT_1}.
In contrast to traditional optimization methods that often become computationally prohibitive in high-dimensional parameter spaces, our approach leverages an incremental-update deep reinforcement learning (IU-DRL) framework to jointly optimize pulse profiles and phases for both control and target atoms.
This synchronous optimization enables the IU-DRL to navigate complex non-adiabatic pathways that often remain inaccessible to conventional piecewise protocols \cite{DRL_non_adiabatic_pathway}. Notably, within this framework, the control atom is no longer required to persistently reside in the Rydberg state during target-atom manipulation. By suppressing redundant Rydberg-state populations and leveraging fast non-adiabatic dynamics, the DRL agent autonomously identifies a control strategy that significantly enhances gate performance and mitigates decoherence.
More importantly, the IU-DRL framework can disclose unexpected control strategies hidden from traditional optimization methods, which are typically limited by fixed horizon control and search policies \cite{porotti2022deep}. As the training proceeds, the agent progressively learns an active early-cutoff strategy, achieving a win-win outcome for both gate duration and gate fidelity. This proves that the IU-DRL is not merely a tool for parameter tuning, but an intelligent paradigm for discovering optimal quantum control pathways.

Numerical simulations, accounting for finite excited- and Rydberg-state lifetimes as well as atomic motion, demonstrate that our protocol compresses the CNOT gate duration by 38.9\% compared to conventional piecewise protocols. 
In the absence of thermal motion, the synchronous-optimized pulse achieves an average fidelity of $0.9991$, significantly surpassing the characteristic $0.99$ fault-tolerant threshold required for surface-code error correction \cite{surface_code_2}. Even at a realistic temperature of 10 $\mu$K, the fidelity remains as high as $0.9988$, maintaining a performance level that consistently challenges prior optimization strategies \cite{high_fidelity_2,high_fidelity_1,high_fidelity_3}.


It is worth noting that while these performance evaluations are conducted within a Master-equation-based simulation environment, the IU-DRL framework is inherently model-free \cite{DRL_model_free_3}. This crucial characteristic implies that the algorithm does not strictly depend on a prior knowledge of the system’s Hamiltonian or noise models. Consequently, our approach is directly extensible to real-world experimental setups, where the agent can learn and evolve by interacting with raw measurement data in a closed-loop fashion. This flexibility positions the IU-DRL as a robust candidate for autonomous gate calibration and cross-talk suppression in complex quantum processors \cite{DRL_model_free_2,suppress_crosstalk_error,quantum_error_correction}.

\begin{figure*}[htbp]
	
	\includegraphics[width=16.5cm]{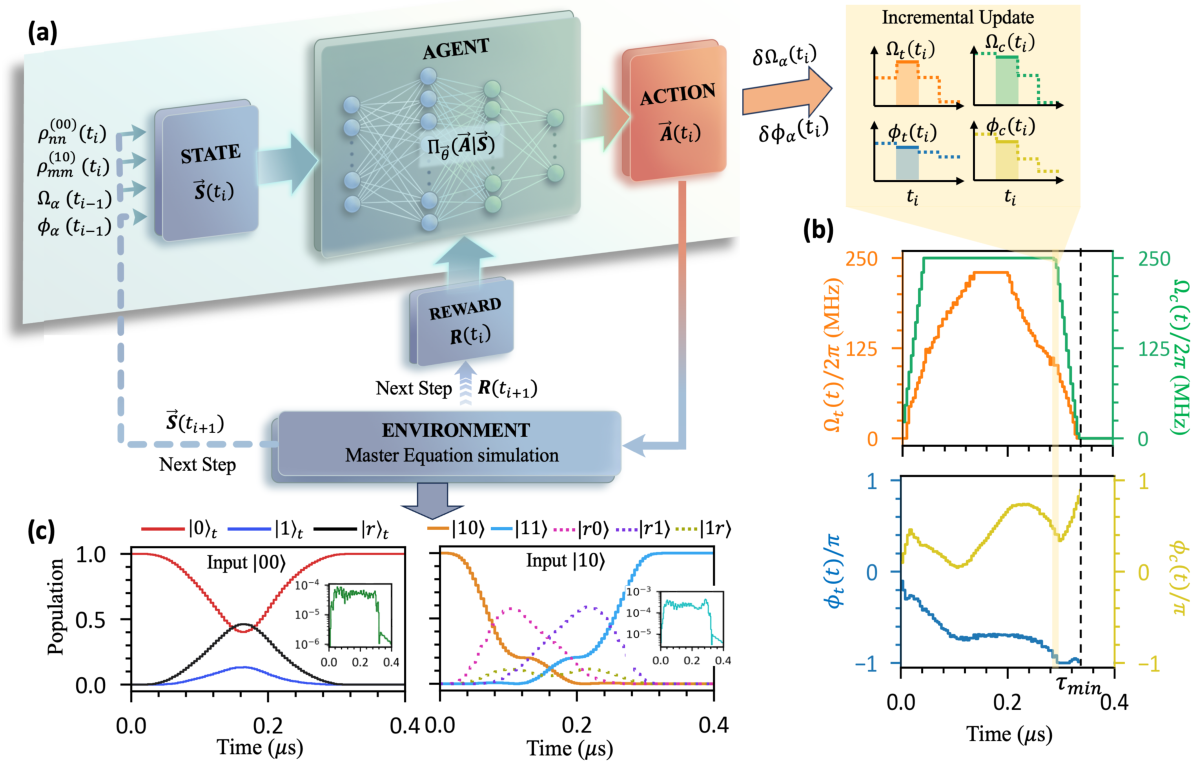} 
	\centering
	\caption{Implementation of Rydberg gates via the IU-DRL intelligent optimization. (a) In the DRL cycle, agent receives the state information $\vec{\mathcal{S}}(t_i)$ from environment at every time step $t_i$ and determines the action $\vec{\mathcal{A}}(t_i)$ according to the learned policy. Meanwhile, the policy is iteratively updated under the guidance of reward signal $\mathcal{R}(t_i)$. Then the environment evolves over the next time step based on the selected action, yielding updated state $\vec{\mathcal{S}}(t_{i+1})$ and updated reward $\mathcal{R}(t_{i+1})$, which are fed back to the agent as inputs for the  $t_{i+1}$ time step. (b) The synchronously-optimized pulse amplitudes $(\Omega_c(t),\Omega_t(t))$ and phases $(\phi_c(t),\phi_t(t))$ for a given duration $t_{N} = 0.4 \mu$s with $N=100$, which are obtained after training for 25,000 epochs, denoted as Case I in Table \ref{tab1}. Depending on the early-cutoff policy, 
    the minimal gate time autonomously discovered by DRL agent is $\tau_{\text{min}}\approx 0.336\mu$s, at which the amplitudes $\Omega_{c}$ and $\Omega_t$ are both turned off.
    Upper Insets: A zoom-in view shows the details of optimized pulses at the $i$th time step using {incremental update} method which are related to the values obtained at $t_{i-1}$ step. (c) The corresponding population evolution for the input states $|00\rangle$ and $|10\rangle$ in which two insets present the population dynamics on state $|e\rangle_t$ (dark green) and sum of all states associated to $|e\rangle_{c,t}$ (light green).
    We set $(\Omega_c^{\max},\Omega_t^{\max})/2\pi = (250,230)$ MHz ensuring a prior excitation of the control atom. The smoothing coefficient is $(\xi_\Omega ,\xi_\phi) = 0.1$. Other parameters are $\Omega_{gl}/2\pi=250$ MHz, $\Delta/2\pi = 7.3$ GHz, $V_0/2\pi=450$ MHz, $\eta_e = 5000,\eta_r = 2000$.}
	\label{DRLfra}
\end{figure*}

\section{REINFORCEMENT LEARNING For CNOT gates}

\subsection{Markov Decision Process for QOC}

The realization of high-fidelity quantum gates with suppressed error rates remains a cornerstone of scalable quantum computing \cite{PRL.135.060602}. Achieving such performance necessitates QOC to precisely modulate pulse amplitudes and phases. However, traditional QOC methodologies often grapple with the ``curse of dimensionality" in complex control spaces, a heavy reliance on accurate gradient information, and inherent vulnerability to systematic uncertainties \cite{QOC_gradient_dependence,QOC_review}.

DRL emerges as a superior paradigm for these challenges, offering distinct advantages over conventional deep learning (DL) approaches. While standard DL typically functions as supervised learning, requiring a massive dataset of pre-existing optimal solutions (labels) for training, DRL is inherently an explorative, label-free framework \cite{RL_book}. In the context of quantum gate design where the ``optimal" pulse shape is often unknown and non-intuitive, DRL enables the autonomous discovery of strategies by interacting directly with the quantum environment. By establishing a closed-loop feedback system, the DRL agent learns through iterative trial and error, uncovering sophisticated control pathways that transcend both human intuition and the limitations of gradient-based optimization.

As illustrated in Fig. \ref{DRLfra}(a), we reformulate the quantum control problem as a finite Markov Decision Process \cite{Markov_1,Markov_2}, where the agent is represented by a multilayer artificial neural network \cite{neural_network}.
The total manipulation duration $T$ is discretized into $N$ equal time steps of length $\delta t$. At each interval $t \in [t_i, t_{i+1}]$, the agent observes the current quantum system state $\vec{\mathcal{S}}(t_i)$ and, guided by its learned policy $\Pi_{\vec{\theta}}(\vec{\mathcal{A}} \mid \vec{\mathcal{S}})$, executes an action vector $\vec{\mathcal{A}}(t_i)$ representing the set of control parameters. To enable the autonomous discovery of optimal strategies, this policy is modeled as a stochastic process where actions are sampled from a multivariate Gaussian distribution
\begin{equation}\vec{\mathcal{A}}(t_i) \sim \Pi_{\vec{\theta}}( \cdot \mid \vec{\mathcal{S}}(t_i) ) = \mathcal{N}\left(\vec{\mu}_{\vec{\theta}}(\vec{\mathcal{S}}), \mathrm{diag}(\vec{\sigma}_{\vec{\theta}}^2)\right).
\end{equation}

This stochastic sampling mechanism is fundamental to the agent’s ability to maintain a balance between exploitation of known high-performance control sequences and exploration of novel trajectories. While the mean vector $\vec{\mu}_{\vec{\theta}}$ represents the agent's current best estimate for the optimal control, the variance $\vec{\sigma}_{\vec{\theta}}^2$ allows the agent to deviate and ``test" alternative control variations. Such exploration is crucial for navigating complex, non-adiabatic landscapes, as it permits the agent to discover non-intuitive control sequences that might be inaccessible to deterministic optimization methods.

Through this interactive framework, the policy parameters $\vec{\theta}$ are iteratively updated to maximize the cumulative expected reward $\mathcal{R}(t_i)$, which is directly coupled to the final gate performance. This optimization process effectively refines the policy, concentrating on the probability mass around action sequences that yield optimal state-evolution trajectories.

\subsection{From Piecewise EIT to Synchronous DRL}

\begin{figure}[htbp]
	\includegraphics[width=6.0cm]{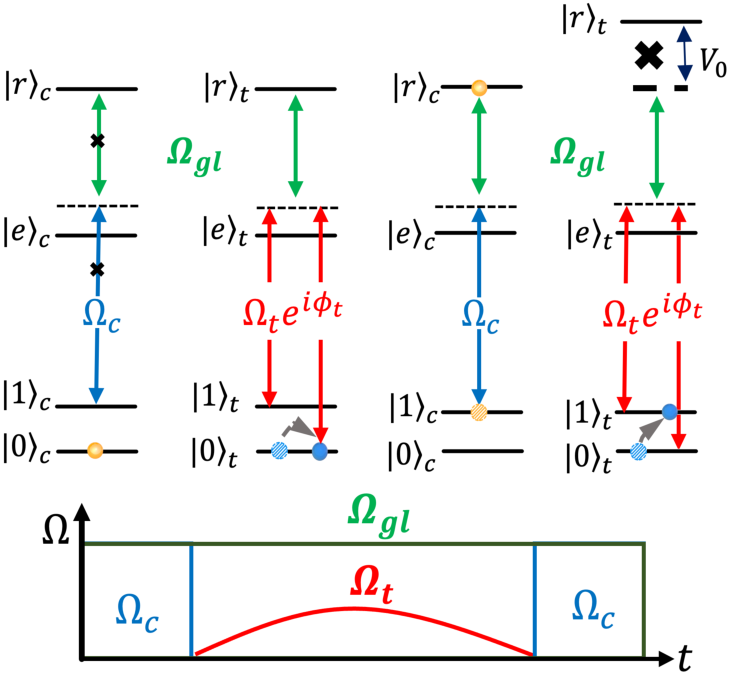} 
	\centering
	\caption{EIT-based piecewise adiabatic gate protocol. STAY: If the control atom is in $|0\rangle_{c}$ the target atom adiabatically follows the EIT dark state $|\Psi\rangle_t = (1/\sqrt{2})[|d_1\rangle_t-|d_2\rangle_t]$ leaving its initial state $|0\rangle_t$ unchanged, i.e. $|0\rangle_c|0\rangle_{t} \rightarrow |0\rangle_c|0\rangle_{t}$. Note that for $|1\rangle_t$ initially, this dark state is $|\Psi\rangle_t = (1/\sqrt{2})[|d_1\rangle_t+|d_2\rangle_t]$. TRANSFER: If the control atom is initially in $|1\rangle_{c}$ the strong vdWs interaction $V_0$ can shift the Rydberg state $|r\rangle_t$ of the target atom, breaking the EIT condition and enabling resonant transfer of $|1\rangle_c|0\rangle_{t} \leftrightarrow |1\rangle_c|1\rangle_{t}$. Bottom: Schematic of the piecewise pulse sequence (not to scale). 
    }
\label{sme}	
\end{figure}

To establish a baseline for intelligent optimization, we first review the conventional gate protocol based on electromagnetically induced transparency (EIT) on a two-Rydberg-atom platform. This scheme is favored for its experimental demonstration and potential scalability to multiqubit operations \cite{EIT_CNOT_2}. As depicted in Fig.~\ref{sme}, two trapped $^{87}\mathrm{Rb}$ atoms (denoted by the subscripts $c$ for control atom and $t$ for target atom, respectively) are both modeled as four-level systems as in \cite{experimental_parameters}. Separated by $r_{0} \approx 2~\mu\mathrm{m}$, they exhibit a Rydberg interaction strength of van der Waals-type, of $V_{0} = 2\pi \times 450$~MHz. The logical qubit states are encoded in the long-lived hyperfine ground states $|0\rangle = |5S_{1/2}, F=1, m_F=0\rangle$ and $|1\rangle = |5S_{1/2}, F=2, m_F=0\rangle$. The intermediate and Rydberg states are specified as $|e\rangle = |6P_{3/2}\rangle$ and $|r\rangle = |53S_{1/2}, m_j=1/2\rangle$, with finite lifetimes $1/\gamma_{e}$ and $1/\gamma_{r}$, respectively. Two-photon excitation from the qubit states to the Rydberg state is driven by a 420-nm $\sigma^{+}$-polarized light and a 1,013-nm $\sigma^{-}$-polarized light. 
The total Hamiltonian governing the system is $\hat{H} = \hat{H}_{c} \otimes I + I \otimes \hat{H}_{t} + V|rr\rangle \langle rr|$, where the individual atomic Hamiltonians are given by
\begin{eqnarray}
\hat{H}_{c} &=& \frac{\Omega_{gl}}{2} |e \rangle_c \langle r| + \frac{\Omega_c(t) e^{i\phi_c(t)}}{2} |1 \rangle_c \langle e|+ \text{H.c.} - \Delta |e \rangle_c \langle e| , \nonumber\\
\hat{H}_{t} &=& \frac{\Omega_{gl}}{2} |e \rangle_t \langle r| + \frac{\Omega_t(t)e^{i\phi_t(t)}}{2} \left( |0 \rangle_t \langle e| + |1 \rangle_t \langle e| \right) + \text{H.c.} \nonumber\\
&-& \Delta |e \rangle_t \langle e|.\label{eq:Ht}
\end{eqnarray}

The conventional EIT-based gate relies on a temporal piecewise sequence where a global Rydberg laser with a constant Rabi frequency $\Omega_{gl}$ off-resonantly drives the $|e\rangle_{\alpha} \leftrightarrow |r\rangle_{\alpha}$ transition with detuning $-\Delta$. Concurrently, three localized piecewise pulses $\{\Omega_c, \Omega_t, \Omega_c\}$ are sequentially applied to either the control or target atom. In this piecewise limit, the phases $\phi_{c,t}$ are typically fixed to zero. When the control atom is in $|0\rangle_c$, the target atom evolves in the unblocked regime where the EIT condition is satisfied, supporting the dark states
\begin{eqnarray}
|d_1 \rangle_t &=& \frac{1}{\sqrt{2}}( |1 \rangle_t - |0 \rangle_t ), \label{d1} \\
|d_2 \rangle_t &=& \frac{1}{\sqrt{1+x^2}} \left[ \frac{|1 \rangle_t + |0 \rangle_t}{\sqrt{2}} - x e^{-i\phi_t(t)} |r \rangle_t \right], \label{d2}
\end{eqnarray}
where $x = \sqrt{2} \Omega_{t}(t)/\Omega_{gl}$. Provided the adiabatic condition $\dot{\Omega}_t(t)\ll \Omega_{gl}^3 / (4\Delta)$ is met, the target atom follows these dark states to complete the gate operation without excited-state $|e\rangle_t$ population loss. This constitutes the STAY operation, $|0 \rangle_c |j \rangle_t \xrightarrow{\text{EIT}} |0 \rangle_c |j \rangle_t$ for $j \in \{0,1\}$. Conversely, if the control atom is in $|1\rangle_c$, it is pumped to $|r\rangle_c$, and the resulting Rydberg blockade $V \gg \Omega_{gl}^2/(2\Delta)$ breaks the EIT condition. This enables the target atom to undergo a Raman $\pi$-pulse $\int dt \Omega_t(t)^2/2\Delta=\pi$, facilitating the TRANSFER operation, $|1 \rangle_c |0 \rangle_t \leftrightarrow |1 \rangle_c |1 \rangle_t$.

Despite its scalability, this adiabatic piecewise design inherently limits the attainable fidelity. Because the control atom must persistently reside in the lossy Rydberg state $|r\rangle_c$ throughout the entire duration of the target-atom manipulation, even a relatively low Rydberg-state decay rate leads to the accumulation of significant decoherence errors $\varepsilon_{control}> 10^{-3}$. 
\textcolor{black}{These accumulated errors typically prevent the gate fidelity from reaching the high-performance 0.999 regime, even if the basic fault-tolerant requirements are met \cite{surface_code_1}.}
Notably, such a performance ceiling cannot be breached even by applying advanced optimization to the individual pulse segments; as long as the sequential structure remains, the inherent Rydberg-state exposure from the control atom remains a decisive error source (see Appendix \ref{appA}).

To transcend these limitations, we introduce a synchronous-optimized approach that abandons the constraints of adiabatic dark-state evolution and piecewise sequencing. In this framework, the pulse profiles for both control and target atoms, including their amplitudes $\Omega_{c,t}(t)$ and phases $\phi_{c,t}(t)$, are modulated simultaneously. By treating the pulse design as a unified optimization problem governed by the full Hamiltonians in Eqs. (\ref{eq:Ht}), the pulses are orchestrated to minimize Rydberg population while exploiting fast, non-adiabatic pathways.

Such a synchronous scheme significantly expands the control landscape, introducing a high-dimensional parameter space that is often inaccessible to traditional quantum optimal control methods \cite{PRA.92.053411,PRA.94.032306,PRR.6.023231}. Conventional algorithms typically struggle with the complexity of simultaneously optimizing multiple continuous variables across discretized time steps without a pre-defined analytical path. This complexity necessitates the application of DRL. As a model-free and explorative framework, DRL is uniquely capable of navigating this expanded landscape to discover non-intuitive pulse strategies \cite{DRL_5}. Through the specifically designed DRL algorithm shown below, the agent learns to effectively suppress unnecessary Rydberg-state populations, thereby achieving a simultaneous enhancement of both gate duration and fidelity. This dynamic learning process enables the discovery of high-performance control pathways that significantly surpass the capabilities of traditional piecewise protocols.

\subsection{Synchronous IU-DRL for CNOT Gates}\label{SynIUDRL}

 {\it The action space vector---} Implementation of high-fidelity CNOT gates requires an efficient mapping of time-dependent external controls onto the DRL action space. These controls include the laser amplitudes $\Omega_{\alpha}(t)$ and phases $\phi_\alpha(t)$, where the subscript $\alpha \in \{c, t\}$ denotes the lasers on control and target atoms, respectively. In conventional DRL schemes, the action vector $\vec{\mathcal{A}}(t_i)$ at each time step typically consists of the raw control parameters $\Omega_{\alpha}(t_i)$ and $\phi_\alpha(t_i)$ \cite{DRL_CTAP}. However, this direct mapping presents two significant challenges. First, it frequently generates non-smooth or strongly oscillatory pulse shapes because the chosen actions can vary abruptly between consecutive time steps. Such erratic control profiles are difficult to implement in practical experiments due to the finite bandwidth and response time of arbitrary waveform generators and acousto-optic modulators between different physical platforms \cite{PRXQuantum.2.030203,PRL.126.070505,PRApplied.19.014062}. Second, the four-dimensional search space, defined by the product of individual control ranges $[0, \Omega_\alpha^{\max}]^2 \times [-\pi, \pi]^2$, substantially increases optimization complexity. This vast landscape not only hinders learning efficiency and slows convergence but also renders the optimization more prone to becoming trapped in suboptimal, local-minima solutions (see Sec. \ref{trf}) rather than identifying the global optimum.

To address these issues, we reformulate the action vector as an incremental update (IU), a framework we refer to as IU-DRL:
\begin{equation}\label{eq:incremental}\vec{\mathcal{A}}(t_i) = \left[\delta \Omega_{c}(t_i), \delta \Omega_{t}(t_i), \delta \phi_{c}(t_i), \delta \phi_{t}(t_i) \right],
\end{equation}
where $\delta \Omega_{\alpha}$ and $\delta \phi_{\alpha}$ denote the changes relative to the values at the previous time step. 
The actual physical control parameters are subsequently obtained via cumulative integration: $\Omega_{\alpha}(t_i) = \Omega_{\alpha}(t_{i-1}) + \delta \Omega_{\alpha}(t_i)$ and $\phi_{\alpha}(t_i) = \phi_{\alpha}(t_{i-1}) + \delta \phi_{\alpha}(t_i)$.
To ensure that the resulting pulses vary smoothly, we introduce smoothing coefficients $\xi_\Omega, \xi_\phi \ll 1$ to constrain the update magnitudes, such that $\delta \Omega_{\alpha}/\xi_\Omega \in [-\Omega_{\alpha}^{\max}, \Omega_{\alpha}^{\max}]$ and $\delta \phi_{\alpha}/\xi_\phi \in [-\pi, \pi]$. 
While the action vector's intrinsic dimensionality remains four, this approach significantly reduces the reachable volume of the control landscape at each step. By prioritizing local, continuous exploration over global, discontinuous jumps, IU-DRL effectively regularizes the search space. This strategy prevents the optimization from stagnating in erratic, suboptimal regions and ensures the discovery of smooth, high-precision control sequences suitable for experimental implementation.

{\it The state vector---} The DRL agent perceives the environment, representing the Rydberg gate system with evolution governed by the master equation (see Appendix \ref{massec}), through an observation vector $\vec{\mathcal{S}}(t_i)$. 
To fully characterize the performance of a CNOT gate, the agent must ideally monitor the evolution of the two-qubit system starting from all four computational basis states: $\{|0\rangle_c|0\rangle_t, |0\rangle_c|1\rangle_t, |1\rangle_c|0\rangle_t, |1\rangle_c|1\rangle_t\}$. 
A brute-force observation of the $16 \times 16$ density matrices for the Rydberg atom system (Fig.~\ref{sme}) across all four initial states would yield a 1024-dimensional state vector, imposing a prohibitive computational overhead on the agent's training.
However, we significantly reduce this high-dimensional observation space by leveraging the physical symmetries of the system.
Specifically, the large laser detuning renders the energy splitting between $|0\rangle_t$ and $|1\rangle_t$ negligible, creating their exchange symmetry that allows the gate's global performance to be characterized by only two initial states: $|0\rangle_c|0\rangle_t$ (STAY) and $|1\rangle_c|0\rangle_t$ (TRANSFER). 
For the STAY operation, the control atom remains a spectator in $|0\rangle_c$, requiring only the target atom to be tracked. 
Furthermore, as high-fidelity CNOT operation in this system is primarily characterized by population dynamics, we restrict the observation vector to the diagonal elements of the density matrices, with off-diagonal coherences offering less orthogonal information for the agent's policy updates.
Consequently, the observation vector is streamlined to only 24 dimensions
\begin{equation}\vec{\mathcal{S}}(t_i) = [\rho_{nn}^{(00)}(t_i), \rho_{mm}^{(10)}(t_i), \Omega_{\alpha}(t_{i-1}), \phi_{\alpha}(t_{i-1})],
\end{equation}
where $\rho_{nn}^{(00)}$ ($n \in \{1,\dots,4\}$) denotes the 4 diagonal elements of the target atom's subspace during the STAY operation, $\rho_{mm}^{(10)}$ ($m \in \{1,\dots,16\}$) represents the 16 diagonal elements of the two-qubit system during the TRANSFER operation, and the final 4 components are two laser parameters from the previous step.
This symmetry-based reduction ensures that the agent captures the essential information while maintaining a compact state representation for efficient training.

{\it Reward function---} Here, the design of the reward function is even pivotal, as it defines the optimization objective and guides the agent's convergence by addressing the inherent delayed reward problem \cite{reward_function}. In quantum gate design, the ultimate performance metric represented by the average gate fidelity across the STAY and TRANSFER operations, $\mathcal{F}_{\text{avg}}$ (see Appendix \ref{massec}), is only observable upon the completion of the full pulse sequence at $t_N$. This sparsity of terminal feedback often leads to a credit assignment challenge, where the agent struggles to correlate specific mid-sequence control actions with the final outcome. To mitigate this, we construct a cumulative reward function that provides both sparse terminal guidance and dense physical feedback
\begin{equation}
\mathcal{R}(t_i) = -\log_{10} \left[ 1 - \delta_{i,N}\mathcal{F}_{\text{avg}}(t_N) \right] - P(t_i), \label{eq:reward_def}
\end{equation}
where the first term represents a terminal reward evaluated only at the final time step $t_N$, with $\delta_{i,N}$ being the Kronecker delta. The logarithmic scaling ensures that as $\mathcal{F}_{\text{avg}}$ approaches unity, the reward increases significantly (e.g., $\mathcal{F}_{\text{avg}} = 0.999$ yields a reward value of 3.0), providing a steep gradient essential for high-precision optimization \cite{DRL_GRAPE}.
The second term, $P(t_i) = \eta_e \gamma_e T_e(t_i) + \eta_r \gamma_r T_r(t_i)$, serves as a dense proxy reward in the form of a step-wise penalty to suppress population leakage. Here, $\gamma_{e,r} T_{e,r}(t_i) = \gamma_{e,r} \int_{t_{i-1}}^{t_{i}} \bar{n}_{e,r}(t) dt$ quantifies the spontaneous decay during each interval, where $\bar{n}_{e,r}$ represents the populations of the intermediate and Rydberg states averaged over all four computational basis states. By weighting these penalties with regularization coefficients $\eta_e$ and $\eta_r$, we balance the terminal fidelity objective against transient decay errors, ensuring robust global optimization and stable learning trajectories.

Notably, the inclusion of this dense penalty term facilitates the discovery of an early-cutoff policy. Unlike traditional piecewise protocols that occupy the Rydberg state for a fixed duration, the IU-DRL agent learns to minimize the time-integrated penalty $P(t_i)$ by turning off laser fields ($\Omega_{c,t} = 0$) at the earliest feasible moment $\tau_{\min} \leq T$ once the target state is reached. This autonomous identification of the optimal truncated duration effectively reduces integrated decoherence, \textcolor{black}{serving as a key factor in reaching the high-fidelity 0.999 regime (see Sec.~\ref{GP}) and significantly surpassing the standard fault-tolerant threshold of 0.99. This approach successfully bypasses the performance bottlenecks inherent in conventional adiabatic designs.}

{\it TRPO algorithm---} Finally, the agent is parameterized via an actor-critic architecture and trained using the Trust Region Policy Optimization (TRPO) algorithm \cite{TRPO}. TRPO is chosen for its ability to ensure monotonic policy improvement through a constraint on the Kullback–Leibler divergence, which is critical for maintaining stability in the high-dimensional, non-convex landscapes of quantum control. 
Both the actor and critic networks employ three fully connected hidden layers with 156, 48, and 16 units, respectively, utilizing hyperbolic tangent activation functions to ensure a smooth, differentiable mapping from the 24-dimensional observation space $\vec{\mathcal{S}}$ to the mean of a 4-dimensional Gaussian action distribution over $\vec{\mathcal{A}}$.
This architecture enables the critic to provide a stable value-function baseline, which effectively reduces gradient variance for the actor’s stochastic Gaussian policy. Consequently, the agent can efficiently explore the continuous control landscape to uncover sophisticated, high-performance synchronous pulses that surpass the limitations of conventional protocols.

\section{Results and Discussions}

\subsection{Gate Performance} \label{GP}

The synchronously-optimized pulses ($\Omega_{c,t}(t), \phi_{c,t}(t)$) synthesized by the IU-DRL agent are presented in Fig.~\ref{DRLfra}(b). Unlike conventional protocols that rely on predefined analytical profiles, our framework generates control sequences through exactly autonomous exploration. The resulting pulses consist of $N=100$ discrete segments, a resolution that balances the smoothness of the control functions with computational efficiency. This discretization effectively casts the QOC problem into a sequential decision-making task, allowing the agent to maximize the cumulative reward defined in Eq.~\eqref{eq:reward_def} through its learned policy.

As shown in Fig.~\ref{DRLfra}(b), the optimized $\Omega_c(t)$ adopts a near-rectangular profile, maintaining the prescribed maximum $\Omega_c^{\max}$ for the majority of the sequence. This amplitude-maximization strategy is physically intuitive: for a complete CNOT operation, the control atom primarily functions as a switch. In the TRANSFER stage, it must undergo an effective $2\pi$ rotation, where minimizing the duration is more advantageous than complex modulation, as shorter gate times directly suppress decoherence error. Conversely, $\Omega_t(t)$ and the associated phases $\phi_{c,t}(t)$ exhibit more intricate modulations. This complexity arises from the dual requirement of accommodating both the STAY and TRANSFER operations while simultaneously accounting for the state-dependent Rydberg blockade and the associated population oscillations.

The corresponding population dynamics are illustrated in Fig.~\ref{DRLfra}(c). When initialized in $|0\rangle_c$, the system evolves through the STAY channel; here, the target atom undergoes non-adiabatic evolution with transient Rydberg-state occupation. In the TRANSFER channel (starting from $|1_c0_t\rangle$), the Rydberg population of the control atom is substantially reduced compared to sequential protocols. This suppression occurs because synchronous modulation removes the requirement for the control atom to remain ``parked" in the Rydberg state throughout the target atom's evolution. Consequently, the dominant Rydberg decay error characterized by $\gamma_rT_r$, typically $\sim 10^{-3}$ in piecewise designs, is suppressed to $3.31 \times 10^{-4}$. This allows the CNOT gate to achieve a total fidelity of $\mathcal{F}_{\text{avg}}\approx0.9991$ with a significantly reduced duration of $0.336\,\mu\text{s}$, representing a speedup of $\sim 38.9\%$ over the optimized non-adiabatic piecewise benchmarks ($\tau_{\text{min}}+t_{\text{sq}}\approx0.550\mu$s, See Table \ref{tab2} in Appendix).

{\it Autonomous early-cutoff---} A key feature of the learned policy is the emergence of early-cutoff behavior. As shown in Fig.~\ref{DRLfra}(b), the agent autonomously terminates the interaction at $\tau_{\min} \approx 0.336\,\mu\text{s}$, which is notably shorter than the prescribed duration $t_N = 0.4\,\mu\text{s}$. The agent discovers that setting $\Omega_{c,t} = 0$ once the target operation is achieved representing the optimal strategy to minimize time-integrated decoherence. This behavior effectively ``freezes" the system at its peak fidelity. 
\textcolor{black}{While conventional QOC methods like GRAPE can be extended to optimize total duration,} they typically operate within a fixed-horizon framework where control segments are treated as independent variables. Consequently, such approaches often utilize the entire allocated duration thus suffer from avoidable fidelity loss. In contrast, IU-DRL formulates the control as a sequential decision-making process where actions are directly coupled to the evolving system state. This allows the policy to naturally suppress further driving upon reaching the target state, realizing an implicit form of time-optimal control \cite{QOC_time_optimal_1}. This autonomous adaptation effectively resolves the fundamental trade-off between gate speed and fidelity, surpassing the limitations of static optimization paradigms \cite{PRA.63.063412}.

\begin{figure}[htbp]
	\includegraphics[width=8.6cm]{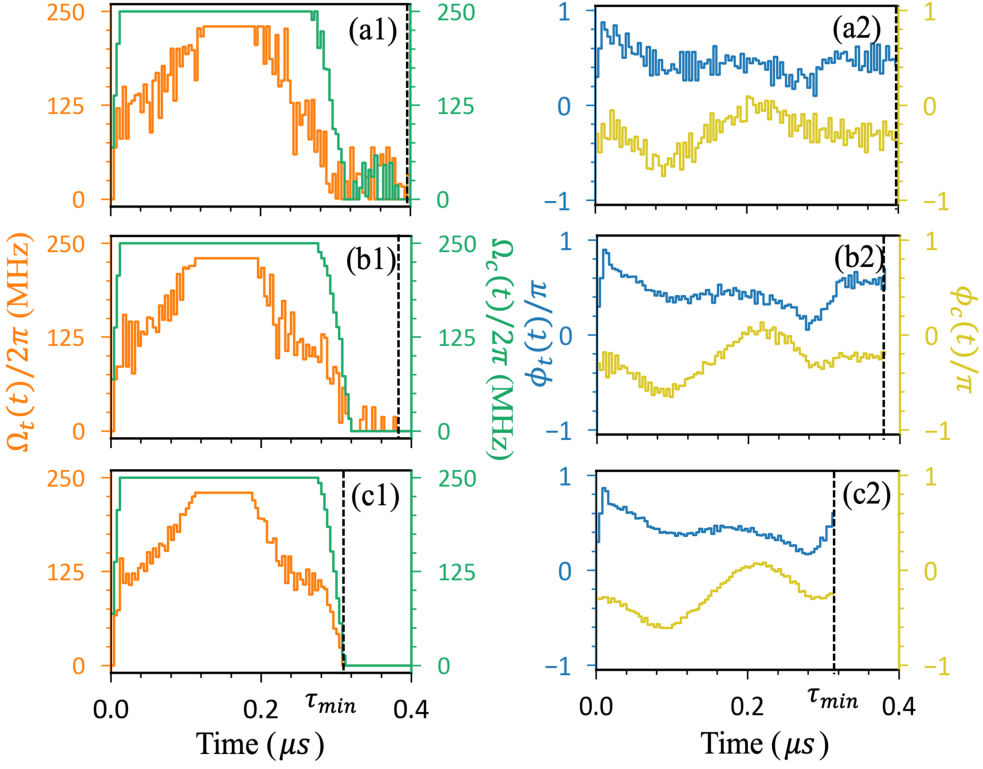} 
	\centering
	\caption{Evolution of IU-DRL-optimized pulse amplitudes and phases across different training stages. The synchronous-pulse protocol follows the setup in Sec.~\ref{GP} but with increased smoothing coefficients $(\xi_\Omega, \xi_\phi) = 0.3$. Panels from top to bottom correspond to Cases II--IV in Table~\ref{tab1}, representing the results after (a1-2) 25,000, (b1-2) 50,000, and (c1-2) 75,000 training epochs, respectively. All other physical parameters are identical to those in Fig.~\ref{DRLfra}b.}
\label{smco}	
\end{figure}

\begin{table*}
	\centering
	\caption{Comparison of gate performance under Incremental Update (IU) and Traditional Update (TU) methods across various training epochs and smoothing coefficients $(\xi_\Omega, \xi_\phi)$. All results are evaluated at the optimal truncated duration $\tau_{\text{min}}$. $\mathcal{F}_{\text{avg}}$ denotes the average fidelity considering spontaneous decay rates ($\gamma_e/2\pi=1$ MHz, $\gamma_r/2\pi=0.5$ kHz), while $\mathcal{F}_{\text{avg}}^r$ incorporates additional decreases $\delta F$ arising from atomic thermal motion at $T=10$ $\mu$K. The decay errors $\gamma_e T_e$ and $\gamma_r T_r$ quantify the integrated population loss in the intermediate and Rydberg states, respectively. For the TU method (Cases V--VIII), each case includes two distinct rows representing the stochastic convergence to different suboptimal strategies (piecewise-like or synchronous-like profiles) observed during training. Notably, TU methods consistently struggle to identify the early-cutoff mechanism, resulting in fixed durations $\tau_{\text{min}} = t_N = 0.4\,\mu$s across all instances.}\label{tab1}
	\begin{tabular}{@{} l c c r 
			S[table-format=1.3] 
			S[table-format=1.5] 
			S[table-format=1.5] 
			S[table-format=1.6]
            S[table-format=1.6] 
			S[table-format=1.6] @{}}
		\toprule
		Method & Case & $(\xi_{\Omega},\xi_\phi)$ & Epoch & {$\tau_{\text{min}}$($\mu$s)} & {$\gamma_{e} T_e$($\times 10^{-4}$)} & {$\gamma_{r} T_r$($\times 10^{-4}$)} & $\mathcal{F}_{\text{avg}}$ & {$\mathcal{F}^{r}_{\text{avg}} (10\mu K)$} & {$\delta F$ ($\times 10^{-4}$)} \\
		\midrule
		\multirow{4}{*}{Incremental update (IU)} 
		& I & 0.1 & 25,000 & 0.336 & 2.72 & 3.31 & 0.99910 & 0.99884 & 2.6 \\
		& II & 0.3 & 25,000 & 0.400 & 6.41 & 3.40 & 0.99194 & 0.99208 & 1.4 \\
		& III & 0.3 & 50,000 & 0.376 & 4.42 & 3.31 & 0.99685 & 0.99715 & 3.3 \\
		& IV & 0.3 & 75,000 & 0.308 & 2.99 & 3.31 & 0.99907 & 0.99883 & 2.4 \\
		\midrule
		\multirow{8}{*}{Traditional update (TU)} 
		& \multirow{2}{*}{V} & \multirow{2}{*}{/} & 25,000 & 0.400 & 116.13 & 4.51 & 0.62720 & 0.63024 & 30.4 \\
		& &  & 25,000 & 0.400 & 112.45 & 5.32 & 0.61793 & 0.62150 & 35.7 \\
		\cmidrule{2-10}
		& \multirow{2}{*}{VI} &  \multirow{2}{*}{/} & 50,000 & 0.400 & 51.20 & 4.37 & 0.82453 & 0.82845 & 39.2  \\
		& &  & 50,000 & 0.400 & 14.17 & 4.26 & 0.91939 & 0.92106 & 16.7 \\
		\cmidrule{2-10}
		& \multirow{2}{*}{VII} &   \multirow{2}{*}{/} & 75,000 & 0.400 & 8.08 & 4.90 & 0.96241 & 0.96163 & 7.8 \\
		& &  & 75,000 & 0.400 & 7.49 & 4.65 & 0.98749 & 0.98669 & 8.0 \\
		\cmidrule{2-10}
		& \multirow{2}{*}{VIII} &  \multirow{2}{*}{/} & 100,000 & 0.400 & 6.76 & 5.26 & 0.99701 & 0.99693 & 0.8 \\
		& &  & 100,000 & 0.400 & 6.97 & 4.94 & 0.99800 & 0.99777 & 2.3 \\
		\bottomrule
	\end{tabular}
\end{table*}

{\it Smoothing coefficients---} In addition, the flexibility of pulse generation is governed by the smoothing coefficients $(\xi_\Omega, \xi_\phi)$, which determine the reachable volume of the control landscape at each time step. We find that $\xi_{\Omega,\phi} = 0.1$ (Case I, corresponding to Fig.~\ref{DRLfra}b) provides an optimal balance between pulse flexibility and stability. Smaller coefficients restrict the agent to suboptimal, adiabatic-like trajectories, whereas larger coefficients expand the action space and may induce pulse oscillations if training is insufficient.
As illustrated in Fig.~\ref{smco}, with $\xi_{\Omega,\phi} = 0.3$ and 25,000 training epochs (Case II), the fidelity reaches only $0.99194$ due to excessive exploration noise within the broader action space. However, by extending the training to 75,000 epochs (Case IV), the pulses converge to smooth profiles with high fidelity ($\mathcal{F}_{\text{avg}} \approx 0.99907$) and a further reduced duration of $\tau_{\min} \approx 0.308\,\mu\text{s}$.

Figure~\ref{smco} also visualizes the progressive learning of the early-cutoff strategy. In the early stages of training (e.g., 25,000 epochs), the policy spans the entire duration $t_N$, though a tendency toward termination begins to emerge. As training reaches 50,000 epochs, the agent first masters the deactivation of $\Omega_c(t)$, which governs the simpler $2\pi$ rotation. Upon reaching 75,000 epochs, both Rabi frequencies converge to the early-cutoff pattern, successfully identifying the minimally accessible gate time under the given physical constraints. In Table~\ref{tab1} we have provided a comprehensive summary of the gate performance across various smoothing coefficients and training epochs. This summary includes a brief categorization of the IU-DRL (Cases I–IV) and Traditional Update (TU) DRL (Cases V–VIII) approaches, which will be discussed in detail later.

\subsection{Comparison with Traditional DRL Framework} \label{trf}

\begin{figure*}[htbp]
	\includegraphics[width=16cm]{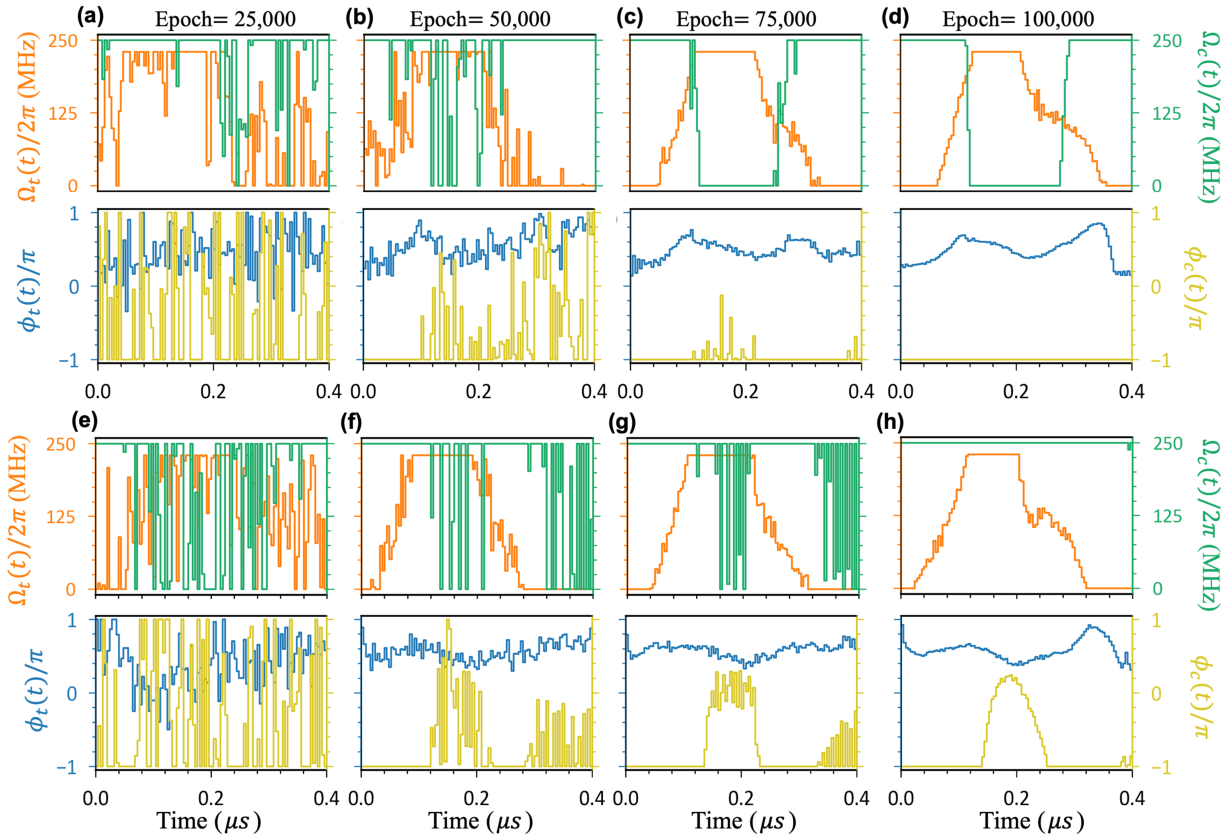} 
	\centering
	\caption{Suboptimal control strategies identified by the DRL agent utilizing the traditional update (TU) method. Panels (a)–(d) and (e)–(h) illustrate two distinct classes of solutions that emerge stochastically as the number of training epochs increases. The first class [(a)–(d)] is analogous to a piecewise-EIT protocol, characterized by the deactivation of the control-atom laser ($\Omega_c(t) \approx 0$) during the target-atom operation. The second class [(e)–(h)] exhibits a synchronous-like profile where $\Omega_c(t)$ remains near its maximum throughout the sequence, yet fails to incorporate the refined early-cutoff mechanism. Detailed gate parameters for these cases are summarized in Table~\ref{tab1} (Cases V–VIII). In both scenarios, the TU method yields lower gate fidelities and longer durations compared to the IU strategy, demonstrating its susceptibility to stagnating in suboptimal solution regions of the high-dimensional control landscape despite significant computational expenditure.
    }
	\label{trad}
\end{figure*}

To evaluate the specific advantages of our IU strategy, we compare its performance against a conventional DRL implementation, hereafter referred to as the Traditional Update (TU) method.
In this scheme, the instantaneous pulse parameters directly constitute the agent's action vector:
\begin{equation}
\vec{\mathcal{A}}(t_i) = [\Omega_c(t_i), \Omega_t(t_i), \phi_c(t_i), \phi_t(t_i)].
\end{equation}
Under this formulation, the agent must explore an expansive, four-dimensional action space at each time step, where $\Omega_{c,t} \in [0, \Omega_{c,t}^{\max}]$ and $\phi_{c,t} \in [-\pi, \pi]$. Unlike the IU approach, the actions in the TU framework are independent of their preceding values, leading to a redundant and discontinuous control landscape \cite{DRL_CTAP,DRL_GRAPE,RL_1}.

Figure~\ref{trad} illustrates the synchronously optimized pulses obtained via the TU-DRL. A primary observation is that the learned pulses exhibit significant stochastic oscillations, with the achievable gate fidelity suffering a drastic reduction ($\mathcal{F}_{\text{avg}} \approx 0.6$) after 25,000 training epochs (see Fig.\ref{trad}(a) and (e), corresponding to Case V in Table~\ref{tab1}). In contrast, the same training budget is sufficient for the IU-DRL agent to achieve near-optimal performance ($\mathcal{F}_{\text{avg}} \approx 0.9991$, Case I) and produce smooth pulses, as previously shown in Fig.~\ref{DRLfra}(b). Increasing the training epochs to 100,000 (see Fig.\ref{trad}(d) and (h), corresponding to Case VIII) eventually improves the fidelity, yet the results remain consistently inferior to the IU strategy. This performance gap stems from the excessive exploration burden imposed by the unconstrained action space, which renders the optimization prone to stagnation in suboptimal regions of the control landscape.

Specifically, the TU framework tends to converge toward two distinct suboptimal solutions with a high degree of stochasticity as training episodes increase. The first strategy [Figs.~\ref{trad}(a)–(d)] mimics the conventional piecewise-EIT protocol by deactivating $\Omega_c(t)$ during the target-atom operation. While this ``piecewise-like" behavior is relatively accessible to the agent due to the simplified profiles of $\Omega_c$ and $\phi_c$, it remains trapped by the inherent fidelity limits of sequential evolution. The second strategy [Figs.~\ref{trad}(e)–(h)] more closely resembles the synchronous IU-DRL results, maintaining a near-maximum $\Omega_c(t)$ throughout the sequence. However, even with 100,000 epochs, the TU agent fails to discover the refined early-cutoff policy. The inability to precisely calibrate the timing of pulse termination under the TU scheme leads to unnecessary Rydberg-state exposure and integrated decoherence, resulting in prolonged gate durations and diminished fidelities compared to the IU-DRL framework. These results, summarized in Table~\ref{tab1}, explicitly demonstrate that the IU-DRL approach effectively regularizes the optimization process. By prioritizing local, continuous updates, the IU-DRL agent efficiently navigates the control landscape to uncover sophisticated pathways that remain inaccessible to traditional DRL methods.

\subsection{Robustness to Atomic Thermal Motion}

\begin{figure}[htbp]
	\includegraphics[width=8.6cm]{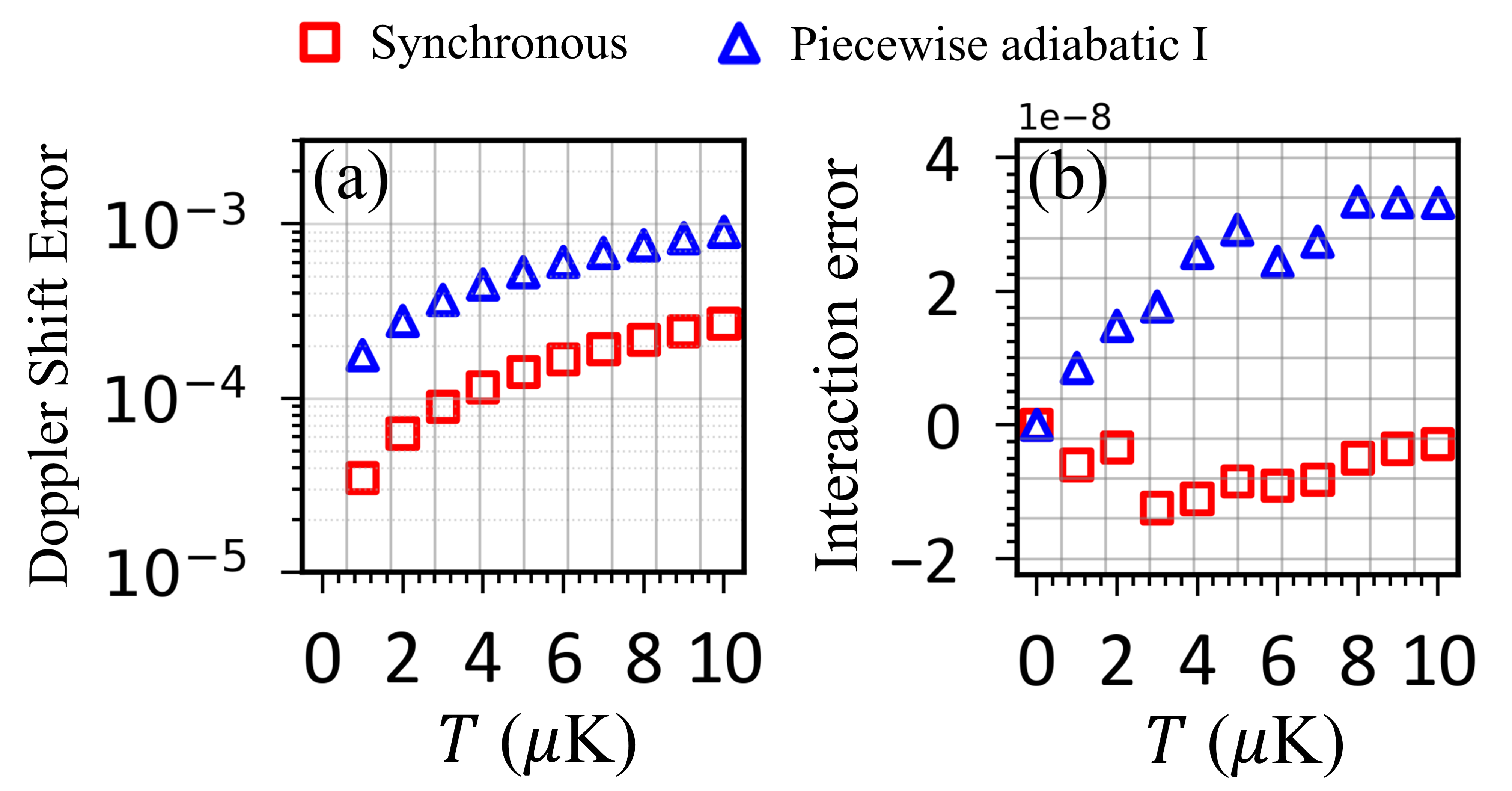} 
	\centering
	\caption{
    Impact of atomic thermal motion on gate infidelity $\delta F$ for the synchronously-optimized IU-DRL protocol I (red squares) and the traditional piecewise adiabatic scheme I (blue triangles). The total infidelity is decomposed into (a) Doppler shift error from random thermal velocities and (b) Rydberg interatomic interaction error from position fluctuations in the optical trap ($\omega/2\pi = 100$ kHz). The synchronous protocol exhibits superior robustness against Doppler-induced dephasing due to its significantly shorter gate duration. In contrast, the interaction error in (b) remains negligible ($< 10^{-7}$) for both schemes, as the gate operates deep within the strong Rydberg blockade regime.}
\label{errors}	
\end{figure}

To evaluate the experimental viability of the IU-DRL optimized protocols, we perform a posterior robustness analysis by subjecting the learned control pulses to a system model that incorporates atomic thermal motion. It is important to emphasize that these thermal effects were not included during the training stage; thus, the following results serve as a stringent test of the intrinsic robustness of the learned control policies against realistic experimental noises \cite{experimental_noise_1,doppler_shift,experimental_noise_2}.

Thermal motion at a finite temperature $T$ introduces two primary dephasing mechanisms. First, the finite velocity of the atoms leads to a stochastic shift in the effective laser frequency, known as the Doppler shift \cite{doppler_shift_Li}. In the presence of this effect, the single-atom Hamiltonian is modified as
\begin{equation}\hat{H}_{\alpha} \to \hat{H}_{\alpha} + \delta_D |r \rangle_{\alpha} \langle r|, \label{eq:doppler_ham}
\end{equation}
where $\delta_D = (k_1 - k_2)v$ is the shifted detuning, with $k_{1,2}$ denoting the wavenumbers of the two-photon excitation lasers. The random velocity component $v$ along the laser propagation direction follows a Maxwell-Boltzmann distribution, which can be treated as a Gaussian distribution with zero mean and standard deviation $\sigma_{v} = \sqrt{k_B T / m}$, where $k_B$ is the Boltzmann constant and $m$ is the atomic mass. To efficiently estimate this temperature impact, we approximate the Doppler shift using the root-mean-square velocity $v_{\text{rms}} = \sqrt{k_B T / m}$ in the calculation.

Figure~\ref{errors}(a) illustrates the gate infidelity $\delta F$ as a function of temperature considering only the Doppler shift. Our numerical results indicate that for the synchronous IU-DRL protocol, $\delta F$ increases only slightly, from $3.49 \times 10^{-5}$ at $T = 1\,\mu\text{K}$ to $2.66 \times 10^{-4}$ at $T = 10\,\mu\text{K}$ (see Table~\ref{tab1}, Case I). This remarkable resilience suggests that the DRL-optimized pulses are intrinsically desensitized to stochastic frequency fluctuations. In contrast, the Doppler-induced error in the traditional piecewise adiabatic case is explicitly worse, reaching nearly $10^{-3}$ at $T = 10\,\mu\text{K}$. This degradation renders the conventional protocol inherently incapable of maintaining the $0.999$ fidelity threshold under realistic experimental conditions \cite{PRL.123.230501}. The superior robustness of the IU-DRL scheme is primarily attributed to the early-cutoff strategy, which substantially compresses the gate duration. By reducing the interaction time, the system effectively minimizes the accumulation of phase errors arising from thermal velocity distributions. Furthermore, this residual error could be further suppressed in experiments by employing laser fields with closer wavelengths \cite{doppler_free_1} or other advanced methods {\it e.g.} three-photon transition \cite{doppler_free_2} and state mapping \cite{doppler_free_3}.

Second, thermal motion also induces the uncertainty of the spacing between two atoms in the trap, leading to fluctuations in the interatomic Rydberg interaction strength $V_0$ \cite{rydberg_interaction_fluctuation}. We model these position fluctuations along the interatomic axis ({\it e.g.} the $\hat{x}$ axis) as a Gaussian distribution with standard deviation $\sigma_{x} = \sqrt{k_B T / m \omega^2}$, where $\omega$ is the trap frequency. For two atoms with independent position fluctuations, the standard deviation of their relative separation is $\sigma_{r} = \sqrt{2} \sigma_{x}$. Then within the van der Waals regime, the temperature-corrected interaction can be formulated as $V' = (r_0 / r_{\text{rms}})^6 V_0$, where $r_{\text{rms}} = \sqrt{r_0^2 + \sigma_r^2}$.

As shown in Fig.~\ref{errors}(b), the influence of interaction fluctuations is effectively suppressed for both gate schemes. This resilience stems from the fact that the protocols operate deep within the strong blockade regime, where the gate fidelity is highly insensitive to the precise value of $V_0$, provided the condition $V_0 \gg \Omega_\alpha^{2}/2\Delta$ is maintained \cite{rydberg_blockade}. Specifically, the gate infidelity arising from interaction fluctuations remains at the $10^{-8}$ level even at $T = 10\,\mu\text{K}$, confirming that Doppler-induced dephasing, rather than interaction noise, is the dominant thermal error source for our Rydberg gates. Nevertheless, a closer inspection reveals that the piecewise adiabatic scheme is still more susceptible to these fluctuations than the IU-DRL protocol. While both deviations are remarkably small, the IU-DRL agent identifies control trajectories that better buffer the system against Rydberg-Rydberg interaction jitter, ensuring superior stability even under subtle environmental perturbations.

\section{Conclusion and outlook}

While Deep Reinforcement Learning (DRL) has demonstrated remarkable potential across various quantum optimal control tasks \cite{DRL_7,DRL_8,DRL_9,DRL_2019NPJ}, its application in synthesizing high-fidelity Rydberg gates remains largely unexplored. In this work, we have theoretically demonstrated the realization of Rydberg CNOT gates utilizing synchronously-modulated control and target pulse profiles, enabled by a quantum optimal control framework based on DRL. Our approach allows the DRL agent to autonomously discover optimal control trajectories without requiring any prior knowledge of pulse symmetry or analytical shapes.

The core of our framework lies in the IU-DRL strategy, which effectively regularizes the exploration of the control landscape by enforcing a tunable smoothing constraint. Unlike traditional DRL methods that often struggle with excessive exploration noise in expansive action spaces, the IU-DRL approach prioritizes local, continuous updates. This ensures the generation of experimentally feasible smooth pulse profiles while simultaneously accelerating convergence. Through this methodology, we have identified a high-performance CNOT gate with an average fidelity of 0.9991 under realistic decay parameters. By autonomously uncovering an early-cutoff strategy, the agent compressed the gate duration to 0.336~$\mu$s, representing a significant improvement over existing protocols in both operational speed and fidelity \cite{neutral_atom_QC_2}. Furthermore, our robustness analysis confirms that these optimized pulses possess an intrinsic resilience against atomic thermal motion, effectively mitigating Doppler-induced dephasing and interatomic interaction fluctuations.

This work establishes a theoretical foundation for employing DRL to design fast, high-fidelity Rydberg logic gates, offering a promising pathway toward large-scale quantum error correction \cite{quantum_error_correction_1,quantum_error_correction_2,quantum_error_correction_3}. Future studies could extend this framework to incorporate noise-adaptive learning during the training stage to achieve active robustness against specific experimental drifts \cite{QOC_and_RL}. Moreover, the IU-DRL approach could be further refined to include hardware-specific constraints, such as limited bandwidth or modulator non-linearities, facilitating the practical translation of these theoretical protocols into real-world neutral-atom architectures.

\section*{Data availability statement}
All data that support the findings of this study are included within the article (and any supplementary files).

\section*{ACKNOWLEDGMENTS}

This work was supported by the Quantum Science and Technology–National Science and Technology Major Project (No. 2021ZD0303200); the National Natural Science Foundation of China (No. 12174106, No. 12374328 and No. 12234014); the Natural Science
Foundation of Chongqing (No. CSTB2024NSCQ-MSX1117); the Science and Technology Innovation Plan of Shanghai Science and Technology Commission (No. 24LZ1400600).

\section*{Conflict of interest}
The authors declare no conflicts of interest.

\section*{ORCID iDs}

\appendix

\widetext
\section{DRL for EIT-based piecewise gates}\label{appA}
In this appendix, we analyze traditional piecewise gate protocols \cite{EIT_CNOT_1}, both adiabatic and non-adiabatic, to provide a benchmark for the synchronous scheme introduced in the main text. To ensure a rigorous and fair comparison, we employ the same IU-DRL framework to optimize the piecewise pulse sequences here, thereby overcoming the limitations of a finite parameter search space inherent in conventional optimization algorithms. The defining feature of the piecewise protocol is its asymmetric control: the Raman pulses $(\Omega_t(t), \phi_t(t))$ for the target atom are optimized, while the control atom is subjected to simple, fixed square $\pi$-pulses $\Omega_c(t)$. The parameters and performance of these cases are listed in Table~\ref{tab2}.

\begin{figure}[htbp]
\includegraphics[width=12.0cm]{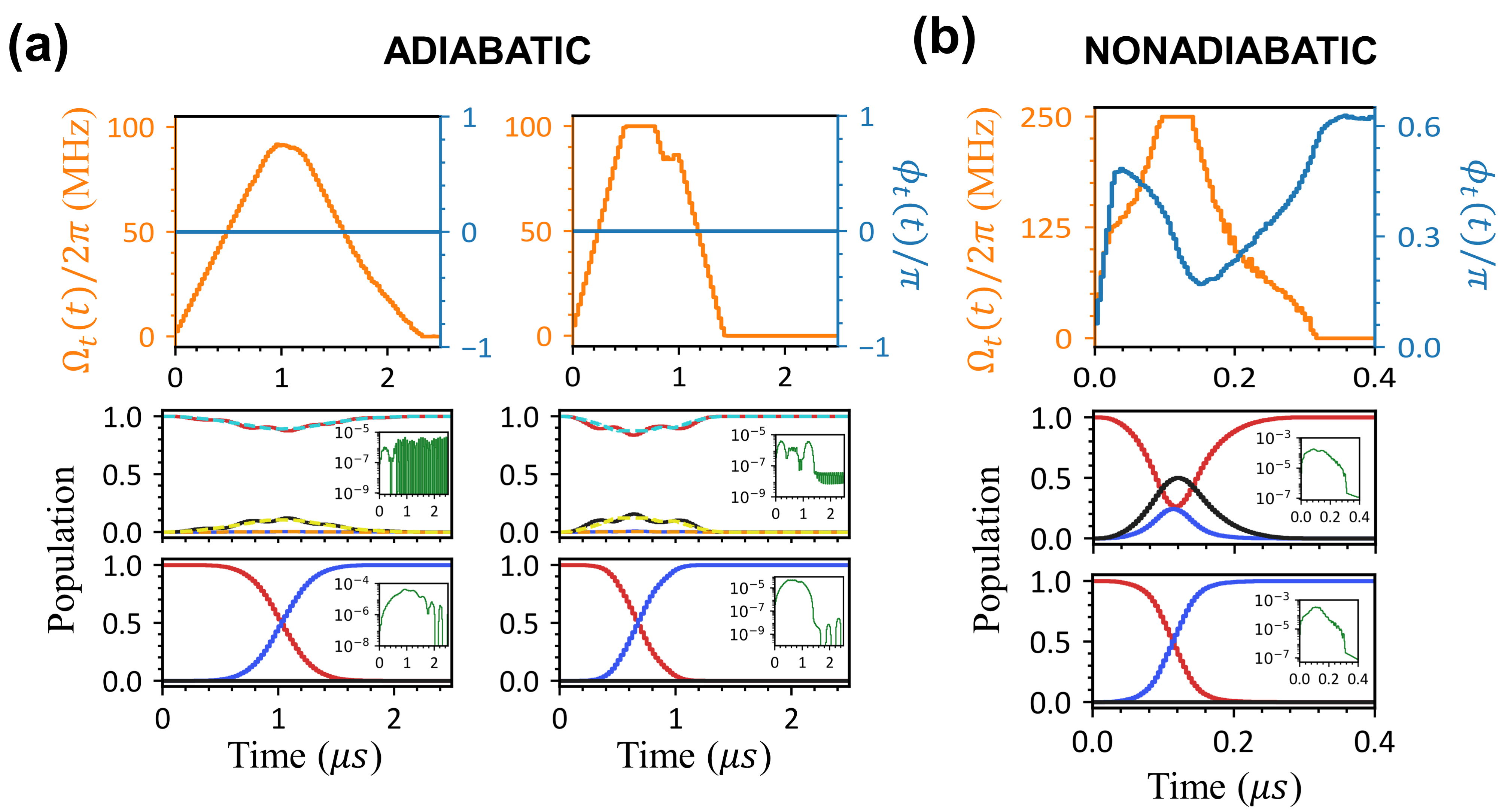}
\centering
\caption{(a) DRL-optimized adiabatic pulse sequences $(\Omega_t(t),\phi_t(t))$ and the resulting population dynamics. The red, blue, and black solid lines represent the populations of states $|0\rangle_t$, $|1\rangle_t$, and $|r\rangle_t$, respectively; insets (green) highlight the suppressed excited-state population $|e\rangle_t$. The analytical results based on the dark state $|\Phi\rangle_t$ are indicated by dashed lines in the middle panels. The left and right columns of (a) correspond to smoothing coefficients of $\xi_\Omega = 0.025$ and $0.05$, respectively, with $\delta\phi_t \equiv 0$ and $\Omega_t^{\max} = \Omega_{gl}/2.5$ to maintain adiabaticity.
(b) Non-adiabatic pulse optimization. By setting $\Omega_t^{\max} = \Omega_{gl}$ and employing larger smoothing coefficients $\xi_\Omega = \xi_\phi = 0.1$, the agent discovers high-speed control trajectories that surpass the EIT regime. In all cases, pulses are discretized into $N=100$ steps and trained over 15,000 epochs. Other parameters are consistent with Fig. \ref{DRLfra}, except for $\Omega_{c}/2\pi = 250$ MHz.
}

\label{piecefig}
\end{figure}

\subsection{Optimized adiabatic Raman pulse}
Under the adiabatic limit, we restrict the optimization to the pulse amplitude $\Omega_t(t)$ while maintaining a constant phase $\phi_t(t) \equiv 0$, as the dark-state population is independent of phase variations in the EIT regime (see Eqs. \ref{d1} and \ref{d2}). The DRL action vector is then defined as $\vec{\mathcal{A}}(t_i) = \delta \Omega_t(t_i)$, where the action space is constrained to $\delta \Omega_{t}(t_i)/\xi_\Omega \in  [-\Omega_{t}^{\max}, \Omega_{t}^{\max}]$ at each time step. To enforce adiabaticity, we utilize a small smoothing coefficient $\xi_\Omega$, ensuring the condition $\dot{\Omega}_t(t) \ll \Omega_{gl}^3 / (4\Delta)$ is satisfied.

In this scheme, the optimization is effectively mapped into the target-atom subspace. A complete CNOT gate is realized if the target atom returns to its initial state (STAY) under the Hamiltonian $\hat{H}_{t}$, but undergoes a complete population transfer (TRANSFER) under the Rydberg-blockaded Hamiltonian $\hat{H}'_{t} = \hat{H}_t + V_0|r\rangle_t\langle r|$. Due to the symmetry of the computational basis, we analyze the evolution starting from $|0\rangle_t$ without loss of generality. To represent this dynamics, the DRL state vector is constructed as
\begin{equation}
\vec{\mathcal{S}}(t_i) = \left[\rho_{nn}^{(00)}(t_i), \rho^{(10)}_{nn}(t_i) \text{ for }n\in\{1,2,3,4\} , \Omega_t(t_{i-1}) \right],
\end{equation}
where $\rho_{nn}^{(00)}$ and $\rho_{nn}^{(10)}$ denote the diagonal elements of the target-atom density matrix evolving under $\hat{H}_{t}$ and $\hat{H}'_{t}$, respectively. The evolution is governed by the Lindblad master equation [Eq. (\ref{mas})], with spontaneous emission restricted to the target subspace.

Figure \ref{piecefig}(a) presents the DRL-optimized adiabatic pulses and the resulting population dynamics. For $\xi_\Omega = 0.025$, the agent generates a smooth adiabatic pulse at the cost of a prolonged gate duration $\tau_{\text{min}} \approx 2.325\,\mu$s. This duration is determined by the autonomously learned early-cutoff policy, which converges to a point that maximizes $F_{t}(t_N)$ while suppressing the intermediate excited-state population below $10^{-5}$. While the target-atom infidelity is remarkably low ($1-\mathcal{F}_t \approx 4.0 \times 10^{-4}$), the total gate performance is severely bottlenecked by the control atom. The control atom, coupled by two square $\pi$-pulses with an extra duration $t_{sq}$, remains in the Rydberg state $|r\rangle_c$ throughout the long adiabatic interval $\tau_{\text{min}}$. This leads to a dominant control-atom error of $\varepsilon_{control} \sim 17.16 \times 10^{-3}$, capping the average fidelity at a modest $\mathcal{F}_{\text{avg}} \approx 0.9910$.

Even when we increase the smoothing coefficient to $\xi_\Omega = 0.05$ to accelerate the adiabatic process, a fundamental trade-off persists. While a faster pulse ($\tau_{\text{min}} \approx 1.450\,\mu$s) reduces the control-atom idling error to $\varepsilon_{control} \sim 11.74 \times 10^{-3}$, the degraded adiabaticity increases the target-atom error. Consequently, the average fidelity remains limited to $\mathcal{F}_{\text{avg}} \approx 0.9932$, cannot reach 0.999 due to intrinsic constraints of the piecewise adiabatic protocol.

\subsection{Optimized non-adiabatic Raman pulses}
To overcome the adiabatic limit, we extend the IU-DRL optimization to the non-adiabatic regime by allowing simultaneous modulation of the pulse amplitude $\Omega_t(t)$ and phase $\phi_t(t)$. The state vector and action space are expanded accordingly
\begin{equation}
\vec{\mathcal{S}}(t_i) = \left[\rho^{(00)}_{nn}(t_i), \rho^{(10)}_{nn}(t_i) \text{ for }n\in \{1,2,3,4 \} , \Omega_t(t_{i-1}),\phi_t(t_{i-1}) \right].
\end{equation}
By setting a larger smoothing coefficient $\xi_\Omega=\xi_\phi = 0.1$ and increasing $\Omega_t^{\max}$ to match $\Omega_{gl}$, the system explores control trajectories beyond the EIT paradigm. 

As shown in Fig. \ref{piecefig}(b), this approach significantly compresses the pulse duration to $\tau_{\text{min}} \approx 0.316\,\mu$s while maintaining a low target infidelity ($3.2 \times 10^{-4}$). Although this represents a substantial improvement over the adiabatic case, the control-atom error, even with the reduced idling time, remains the primary limiting factor at $\varepsilon_{control} \sim 4.6 \times 10^{-3}$. The maximum attainable average fidelity is $\mathcal{F}_{\text{avg}} \approx 0.9974$ with a total gate time of $0.550\,\mu$s. These results indicate that the piecewise scheme, owing to its design, leaves the control atom vulnerable to decoherence while the target atom evolves. This inherent asymmetry prevents the gate from reaching the $0.999$ fidelity level, a limitation that is successfully overcome by the synchronous optimization strategy proposed in the main text.

\begin{table}
	\centering
	\caption{Comparison of gate performance for different piecewise protocols optimized by IU-DRL in Appendix \ref{appA}. Here, $\tau_{\text{min}}$ denotes the optimal truncated duration of the Raman pulse applied to target qubit, while $t_{sq}$ corresponds to the duration of the two square $\pi$-pulses applied to control qubit. The total gate time for piecewise protocols is $\tau_{\text{min}} + t_{sq} $. $\mathcal{F}_{t} = ( \mathcal{F}_{00} +  \mathcal{F}_{10} ) /2$ denotes the average target fidelity evaluated by neglecting gate errors of the control qubit. $\varepsilon_{control}$ denotes the error arising from the imperfect $|1\rangle_c \rightarrow |r\rangle_c \rightarrow |1\rangle_c $ process individually for the control qubit.}\label{tab2}
	\begin{tabular}{@{} l c l c
			S[table-format=1.5]
            c
			S[table-format=1.5] 
			S[table-format=1.5]
            S[table-format=1.5] 
			  @{}}
		\toprule
		Protocol & \text{ }Epoch\text{ } & \text{ }$\xi_{\Omega}$ \text{ }& \text{ }$\xi_\phi$\text{ } & \text{ }{$\tau_{\text{min}}$($\mu$s)}\text{ } & \text{ }$t_{sq}$($\mu$s) \text{ }& \text{ }$\mathcal{F}_{t}$\text{ } & \text{ }{$\varepsilon_{control}$($\times 10^{-3}$)} \text{ }& \text{ }$\mathcal{F}_{\text{avg}}$ \\
		\midrule
		Adiabatic I & \multirow{3}{*}{15,000} & 0.025 & / & 2.325 & \multirow{3}{*}{{0.234}} & 0.99960 & 17.16 & 0.99102 \\
		Adiabatic II & & 0.05 & / & 1.450 &  & 0.99911 & 11.74 & 0.99324 \\
		Non-adiabatic & & 0.1 & 0.1 & 0.316 &  & 0.99968 & 4.60 & 0.99738 \\
		\bottomrule
	\end{tabular}
\end{table}

\section{Master equation and fidelity} \label{massec}
In the presence of spontaneous emission, the dynamics of the two-atom system are described by the Lindblad master equation
\begin{equation}
\frac{ d\hat{\rho} } {dt} = -i \left[ \hat{H}, \hat{\rho} \right] + \hat{\mathcal{L}}[\hat{\rho}],\label{mas}
\end{equation}
where the dissipator $\hat{\mathcal{L}}[\hat{\rho}] = \sum_{\alpha\in\{c,t\}} ( \hat{\mathcal{L}}_e^{\alpha} [ \hat{\rho} ] + \hat{\mathcal{L}}_r^{\alpha} [ \hat{\rho} ] )$ accounts for decoherence in both the control ($c$) and target ($t$) atoms. The individual Lindblad terms are given by
\begin{equation}
\label{eq11}\hat{\mathcal{L}}_k^{\alpha} [ \hat{\rho} ] = \sum_{j=0,1} \left( \hat{L}_{kj}^{\alpha} \hat{\rho} \hat{L}_{kj}^{\alpha\dagger} - \frac{1}{2} \{ \hat{L}_{kj}^{\alpha\dagger} \hat{L}_{kj}^{\alpha}, \hat{\rho} \} \right),
\end{equation}
where $k \in \{e, r\}$ and $\{ \cdot, \cdot \}$ denotes the anti-commutator. The jump operators $\hat{L}_{e j}^{\alpha} = \sqrt{\gamma_{e}/2} |j \rangle_{\alpha} \langle e|$ and $\hat{L}_{r j}^{\alpha} = \sqrt{\gamma_{r}/2} |j \rangle_{\alpha} \langle r|$ represent spontaneous decays from the intermediate state $|e\rangle$ and Rydberg state $|r\rangle$ to the computational basis states $|j\rangle$ with rates $\gamma_e$ and $\gamma_r$.

\subsection{Synchronous-pulse protocol}
For the synchronous modulation scheme, we evaluate the system's evolution starting from the computational basis states $\{ |\mu\rangle \}$, where $\mu \in \{00, 01, 10, 11\}$. 

(i) For initial states $|00\rangle$ or $|01\rangle$, the control atom remains in $|0\rangle$ and is effectively decoupled. The dynamics reduce to the target-atom subspace
\begin{equation}
\frac{ d\hat{\rho}^{(\mu)} } {dt} = -i \left[ \hat{H}_{\mu}, \hat{\rho}^{(\mu)} \right] + \hat{\mathcal{L}}_e^{t} [ \hat{\rho}^{(\mu)} ] + \hat{\mathcal{L}}_r^{t} [ \hat{\rho}^{(\mu)} ]. 
\label{massy1}
\end{equation}

(ii) For initial states $|10\rangle$ or $|11\rangle$, both atoms are involved, and the full master equation [Eq. (\ref{mas})] is solved using the respective Hamiltonians $\hat{H}_{10}$ and $\hat{H}_{11}$.

To quantify the gate performance, we define the state fidelity for each input $|\mu\rangle$ as
\begin{equation}
\mathcal{F}_{\mu} = \text{Tr}\left[ \sqrt{ \sqrt{\rho_{id}^{(\mu)}} \rho_{com}^{(\mu)}(t_N) \sqrt{\rho_{id}^{(\mu)}} } \right]^2,
\end{equation}
where $\rho_{id}^{(\mu)}$ is the density matrix of the ideal CNOT output for input $|\mu\rangle$, and $\rho_{com}^{(\mu)}$ is the evolved density matrix projected onto the computational subspace $\{|00\rangle, |01\rangle, |10\rangle, |11\rangle\}$. The average gate fidelity is then given by $\mathcal{F}_{\text{avg}} = \frac{1}{4} \sum_{\mu} \mathcal{F}_{\mu}$.

\subsection{Piecewise-pulse protocol}
In the piecewise scheme, the control atom is independently driven by a square $\pi$-pulse sequence. The dynamics can be simplified by analyzing the target atom's reduced density matrix $\hat{\rho}^{(00)}$ (for control in $|0\rangle$) and $\hat{\rho}^{(10)}$ (for control in $|1\rangle$). 
The master equations are
\begin{equation}
\frac{ d\hat{\rho}^{(\nu)} } {dt} = -i \left[ \hat{H}_{\nu}, \hat{\rho}^{(\nu)} \right] + \hat{\mathcal{L}}_e^{t} [ \hat{\rho}^{(\nu)} ] + \hat{\mathcal{L}}_r^{t} [ \hat{\rho}^{(\nu)} ],
\end{equation}
where $\nu \in \{00, 10\}$ and $\hat{H}_{00}=\hat{H}_t$, $\hat{H}_{10}=\hat{H}'_t$. The corresponding state fidelities $\mathcal{F}_{00}$ and $\mathcal{F}_{10}$ are calculated relative to the ideal identity and NOT operations on the target qubit, respectively.

To account for the decoherence of the control atom during its $|1\rangle \leftrightarrow |r\rangle$ transition, the average fidelity is corrected as
\begin{equation}
\mathcal{F}_{\text{avg}} = \frac{1}{4} \left[ 2\mathcal{F}_{00} + 2 ( \mathcal{F}_{10} - \varepsilon_{control} ) \right],
\end{equation}
where $\varepsilon_{control}$ represents the gate error incurred by the control atom, primarily due to spontaneous decay while residing in the Rydberg state.

\bibliography{refs}

\end{document}